\documentstyle[amssymb,12pt,thmsa,sw20lart]{article}

\input{tcilatex}
\begin{document}

\title{Gradually Truncated Power Law distribution - citation of scientists }
\author{Hari M. Gupta, Jos\'{e} R. Campanha and B. A. Ferrari \and Unesp - IGCE -
Physics Dpto. \and Rio Claro - S\~{a}o Paulo - Brazil}
\maketitle

\begin{abstract}
The number of times, a scientist is cited in other scientific publications
is now an important factor in his merit consideration. Normally citation
indices of highly cited scientists are available, which in turn give
information only about the dynamics of the citation mechanism of this group.
In the present work, we studied the statistical distribution of the citation
index of Brazilian scientists working in diverse areas, through Zipf-plot
technique. As it is a small sub-group within the scientific community, it
can better explain the dynamics of citation index. We find that gradually
truncated power law distribution can explain well citation index. The
distribution of citation of most cited physicist can also be well explained 

We develop a model of citation, based on positive feedback, i.e. highly
cited scientist, would get better financial help, and more students, which
in turn help to form a large group working in the same topic and paper is
cited more times. The limiting factor comes because of limited number of
articles which can be published in a particular sub-area due to limited
number of scientific journal, where the article can be cited.
\end{abstract}

{\bf I. Introduction}

\bigskip

Power law distribution [1-5] has been first noted by Pareto in economics [6]
in 1897 and afterward by others in many physical [7-10], biological [11-12],
economical [13-17] and educational [18] complex systems. Recently we have
shown that by gradually truncating power law distribution after certain
critical value, we are able to explain the entire distribution including for
very large steps in financial and and physical complex systems [19,20]. We
consider that power law distribution is due to positive feedback which
ceases gradually after certain step size due to limited physical capacity of
the components of the system or system itself.

Scientific publication is a primary means of scholarly communication in
science. The merit of a scientist or a scientific paper is normally
considered through the number of times he or the paper is cited in other
scientific works. Although this is not an exact measurement of the
importance of a paper, but can be considered as a good measurement. While
the average or total number of citations are often quoted anecdotally and
tabulations of highly-cited papers or scientists exist, the focus of this
work is on more fundamental distribution of citations, namely, the number of
scientist in a particular area which have been cited a total of $x$ times, $%
N(x)$.

In a 1957 study based on the publication record of the scientific research
staff at Brookhaven National Laboratory, Shockley [21] claimed that the
scientific publication rate is described by a log-normal distribution.
Laherrere and Sornette [22] have presented numerical evidence, based on data
of the 1120 most-cited physicists from 1981 to June 1997, that the citation
distribution of individual authors has a stretched exponential form, $N(x)$ $%
\sim $ $\exp (-\frac{x}{x_{0}})^{\beta }$ with $\beta \simeq 0.3$. Using the
technique of the Zipf plot, Redner [23] recently showed that the
distribution of citation of a scientific paper is described by a power law $%
N(x)\sim x^{-\alpha }$ with $\alpha \simeq 0.3.$

\smallskip

II. The model

\smallskip

Generally the citation index of highly cited scientist is avalaible. It is
therefore interesting to construct a Zipf plot [24], in which the number of
citations of the $k^{th}$ most ranked scientist out of an ensemble of $M$
scientists is plotted versus rank $k$. By its very definition, the Zipf plot
is closely related to the cumulative large-$x$ tail of the citation
distribution. This plot is therefore well suited for determining the large-$%
x $ tail of the citation distribution. This plot also smooths the
flutuactions in the high-citation tail and thus facilitates quantitative
analysis.

Given an ensemble of M scientists and the corresponding number of citations
for each of these scientists in rank order $Y_{1}\geqslant Y_{2}\geqslant
Y_{3}...\geqslant Y_{n}\geqslant ...Y_{M}$, then the number of citations of
the $n^{th}$ most-cited scientist $Y_{n}$ may be estimated by the criterion
[24]:

\smallskip

\begin{equation}
\int_{Y_{n}}^{\infty }N(x)dx=\int_{Y_{n}}^{\infty }M.P(x)dx=n
\end{equation}

\smallskip

This specifies that there are $n$ scientists out of the ensemble of $M$
which are cited at least $Y_{n}$ times. From the dependence of $Y_{n}$ on $n$
in a Zipf plot, one can test wheter it accoords with a hypothesised form for 
$N(x)$.

We consider that gradually truncated power law distribution also hold good
in case of citation index as in other cases [18-20]. The fact that a
scientist is cited more times facilitate him to get more financial help to
his research projects and better students, which in turn form a better and
larger groups and the paper is initially cited more times in papers of the
same group. Further an more cited article become well known to other
scientist in the same area and is cited by others for completing
introduction of the problem. The more cited scientist can more efficiently
level their citation index than the average scientist, creating more
citation and achieve an higher level of citation index. Thus, the positive
feedback mechanism enhance the production of whatever is being measured.
This feedback effect decreases gradually after certain step size due to
physical limitation of the system. In the present case the limitation came
from limited number of scientific journal and thereby limited number of
scientific articles published in the area in which the scientist can be
cited. This whole process gives Gradually Truncated power law distribution
[18], we can describe as follows:

\smallskip

\begin{equation}
P(x)=\frac{c_{1}}{c_{2}+(\left| x-x_{m}\right| )^{1+\alpha }}f(x)
\end{equation}

\smallskip

\noindent $P(x)$ is the probability of taking a step of size $x$, $x_{m}$ is
the value of $x$ for maximum probability, $c_{1}$ and $c_{2}$ are constants
and are related through:

\smallskip

\begin{equation}
c_{1}=c_{2}P(x_{m})
\end{equation}

\smallskip

\noindent $c_{2}$ can be obtained through normalization condition. Further

\smallskip

\begin{equation}
f(x)=\left\{ 
\begin{array}{ccl}
1 & \mbox{if} & \left| x\right| \leqslant x_{c} \\ 
\exp \left\{ -\left( \frac{\left| x\right| -x_{c})}{k}\right) ^{\beta
}\right\} & \mbox{if} & \left| x\right| >x_{c}
\end{array}
\right.
\end{equation}

\smallskip

\noindent where $x_{c}$ is the critical value of step size where probability
distribution began to deviate from power law distribution due to physical
limitation, $k$ gives the sharpness of the cut-off and $\beta $ is related
to $\alpha $ by:

\smallskip

\begin{equation}
\beta =2-\alpha
\end{equation}

\smallskip

In the present analysis, we are consirering only more cited physicists, so $%
x_{m}$ and $c_{2}$ are almost negligible in compare to $x$ and we neglct
them. Now we consider two special cases:

\smallskip

case $I$: when $x\leqslant x_{C}$

\smallskip

\begin{equation}
N(x)=cx^{-1-\alpha }
\end{equation}

\smallskip

\noindent where $c=c_{1}M$. Through Equation (1), we get:

\smallskip

\begin{equation}
Y_{n}=c_{1}M\alpha n^{-\frac{1}{\alpha }}
\end{equation}

\smallskip

\noindent or

\smallskip

\begin{equation}
\log Y_{n}=(-\frac{1}{\alpha })\log n+b
\end{equation}

\smallskip

i.e. $\log Y_{n}$ vs. $\log n$ will give a straight line as expected in
power law distribution. $b$ is constant.

\smallskip

case II: $x\geqslant x_{C}$

\smallskip

In this case variation due to $f(x)$ is predominant compare to power law and
thus

\smallskip

\begin{equation}
N(x)\sim \exp \left\{ -\left( \frac{\left| x\right| }{k)}\right) ^{\beta
}\right\}
\end{equation}

\smallskip

\noindent or

\smallskip

\begin{equation}
Y_{n}^{\beta }=-\ln n+b
\end{equation}

\smallskip

a and b are constants. This correspond to stretched exponential distribution
[22].

\smallskip

{\bf III. Discussion}

\smallskip

In the present paper, we analyze citation index of (a) most-cited Brazilian
physicists and (b) most cited physicist in whole world. All physicist
including Brazilian physicists publish their work in same Journals and work
almost on same problems due to basic nature of the subject. The physics,
like any other basic science, is same throughout the world. The Brazilian
physicist form a small group within the physicist community. we therefore
assume that power law index ($\alpha $) and limiting step size ($x_{C}$)
must be the same in both cases. However as Brazilian physicist is a small
group, so limiting factor is important only for a few top ranking scientist.
For the rest of the most cited physicist, only power law is important, and
thus index of power law can be well evaluated. In case of internationally
most cited physicist, limiting factor may be important for almost all
physicists.

In Figure (1) we plot citation number versus rank for first 100 Brazilian
physicists in year 1999. We observe a straight line for higher values of $k$
as is expected in case $I$, which begin to deviate for lower values of $%
k(k<20)$. we plot the theoretical curve considering $\alpha =2.53$, $%
x_{C}=2000$ and $k=1500$. The agreement of theoretical curve with empirical
results is good.

In Figure (2), we plot citation numbers versus rank for 1120 most cited
physicist over the period 1981-June 1997 [22] with the same values of
parameter as used in Figure (2) and compare it with the theoretical curve.
We changed the value of constant from $2.10^{6}$ to $1.10^{9}$ because total
number of physicist is much larger in this case. We are considering
brazilian physicist production roughly $0.2\%$ of total physicist production
which is reasonable. We again observe a good agreement. Note that we are
able to explain both distributions with same values of basic parameters. In
present case all the most cited physicist have citation index superior than $%
x_{C}$ and therefore stretched exponential distribution can be considered
for this distribution as is done by Laherrere and Sornette [22]. However we
feel that this distribution is a combination of power law and a stretched
exponential distribution, as also is case of citation index for Brazilian or
any other small sub-group of the whole physicist community. Thus Gradually
Truncated power law distribution can explain the distribution on the whole
scale as we also observed in other fields [18-20]. We feel that the same
must be the case in others basic sciences like chemistry, biochemistry,
mathematics etc.. However in case of social sciences and partially in
biological sciences, local effect may also be important and thus parameters
vary from small sub-group to entire community.

{\bf Figure Captions}

\smallskip

\smallskip

\noindent {\bf Figure 1} Log-Log plot of the dependence of the $n^{th}$ rank 
$Y_{n}$ as a function of rank $n,$ where $Y_{n}$ is the total number of
citations of the $n^{th}$ most cited brazilian physicist. ..... are the
empirical points while straight line is a theoretical curve with gradually
truncated power law distribution.

\smallskip

\smallskip

\noindent {\bf Figure 2} Log-Log plot of the dependence of the $n^{th}$ rank 
$Y_{n}$ as a function of rank $n,$ where $Y_{n}$ is the total number of
citations of the $n^{th}$ most cited physicist. Broken line is empirical
curve while straight line is a theoretical curve with gradually truncated
power law distribution using same parameters as in Figure (1).

\smallskip

\smallskip \newpage

\end{document}